\documentclass[aps,prd,twocolumn,groupedaddress]{revtex4}
\usepackage{amsfonts}
\usepackage{mathrsfs}
\usepackage{amsmath}
\usepackage{amssymb}
\usepackage{bm}
\usepackage{extarrows}
\usepackage{slashed}
\usepackage{isodateo}
\usepackage{xcolor}
\usepackage{graphicx}
\usepackage[CJKbookmarks=true, bookmarksnumbered=true,bookmarksopen=true]{hyperref}
\hypersetup{colorlinks,%
            linkcolor=[rgb]{0,0.2,0.6}, %
            citecolor=[rgb]{0,0.2,0.6}}
\usepackage{indentfirst}

\renewcommand{\FR}[2]{\displaystyle\frac{\,{#1}\,}{#2}}
\newcommand{\fr}[2]{\mbox{$\frac{\,{#1}\,}{#2}$}}
\newcommand{\n}{\nonumber}

\newcommand{\beq}{\begin{equation}}
\newcommand{\eeq}{\end{equation}}
\newcommand{\bq}{\begin{equation}}
\newcommand{\eq}{\end{equation}}
\newcommand{\ba}{\begin{array}}
\newcommand{\ea}{\end{array}}
\newcommand{\beqa}{\begin{eqnarray}}
\newcommand{\eeqa}{\end{eqnarray}}
\newcommand{\beqs}{\begin{subequations}}
\newcommand{\eeqs}{\end{subequations}}
\def\bge{\begin{equation}}
\def\ede{\end{equation}}
\def\bga{\begin{aligned}}
\def\eda{\end{aligned}}
\def\bgp{\begin{pmatrix}}
\def\edp{\end{pmatrix}}
\def\bgs{\begin{subequations}}
\def\eds{\end{subequations}}
\newcommand{\order}[1]{\mathcal{O}({#1})}

\def\nn{\nonumber}

\def\({\left(}
\def\){\right)}

\def\mb{\mathbf}

\def\pd{\partial}
\def\ld{{\mathscr{L}}}

\def\la{\langle}\def\ra{\rangle}

\def\tr{\mathrm{\,tr\,}}

\def\ii{\mathrm{i}}

\def\be{\beta}

\def\de{\delta}
\def\ep{\epsilon}

\def\lam{\lambda}

\def\X{\chi}

\def\ZZ{\mathbb{Z}_2^{}}

\def\Mp{M_{\text{Pl}}^{}}

\def\End{\end{document}}

\begin{document}

\title{\large New Higgs Inflation in a No-Scale Supersymmetric SU(5) GUT}

\author{{John\ Ellis}\,$^{a}$\footnote{Email: John.Ellis@cern.ch},~
        {Hong-Jian He}\,$^{b}$\footnote{Email: hjhe@tsinghua.edu.cn},~
        {Zhong-Zhi Xianyu}\,$^b$\footnote{Email: xianyuzhongzhi@gmail.com}}
\affiliation{$^a$Theoretical Particle Physics and Cosmology Group, Department of Physics,
                King's College London, London WC2R 2LS, UK; \\
                Theory Division, CERN, CH-1211 Geneva 23, Switzerland. \\
             $^b$Institute of Modern Physics and Center for High Energy Physics,
                 Tsinghua University, Beijing 100084, China; \\
                 Center for High Energy Physics, Peking University, Beijing 100871, China.
             }

\begin{abstract}
Higgs inflation is attractive because it identifies inflaton with the electroweak Higgs boson.
In this work, we construct a new class of supersymmetric Higgs inflationary models in no-scale
supergravity with an SU(5) GUT group. Extending the no-scale K\"ahler potential and SU(5) GUT
superpotential, we derive a generic potential for Higgs inflation that includes the quadratic
monomial potential and a Starobinsky-type potential as special limits. This type of models can
accommodate a wide range of the tensor-to-scalar ratio $\,r=\order{10^{-3}\!-10^{-1}}$\,,\,
as well as a scalar spectral index $\,n_s^{} \sim 0.96$\,.
\\[1.5mm]
PACS numbers: 98.80.Cq, 04.65.+e, 12.10.-g, 12.60.Jv
\\[1mm]
{Phys.\ Rev.\ D}\,--\,{Rapid Communication [\,arXiv:1411.5537\,].}
\\[1mm]
{\tt KCL-PH-TH/2014-48, LCTS/2014-49, CERN-PH-TH/2014-230}
\end{abstract}

%

\maketitle


\noindent
{\bf 1.~Introduction}
\vspace*{3mm}

Cosmological inflation \cite{inflation} resolves conceptual dilemma of the standard big-bang
cosmology such as the horizon and flatness problems, and predicts that large-scale structures
in the Universe originated from a nearly scale-invariant spectrum of density perturbations,
which is well consistent with cosmological observations~\cite{cmb1,cmb2}.
Theories of cosmological inflation postulate a scalar field, the inflaton, whose
field energy drove an early epoch of near-exponential expansion.
Before the LHC Higgs discovery, it was very tempting to identify this inflaton as the Higgs boson
of the Standard Model (SM)~\cite{hi}, a very economical and predictive scenario.
However, the recent LHC and Tevatron measurements of the Higgs and top quark masses
indicate that the SM Higgs potential turns negative at $\sim 10^{11}$\,GeV~\cite{INS},
which is lower than the typical cosmic inflation scale $\sim 10^{16}$\,GeV.
This means that new physics is indispensable for our universe to reach
a stable electroweak vacuum after inflation.

\vspace*{1mm}

On the other hand, the SM suffers
from the naturalness problem of stabilizing the electroweak scale against radiative
corrections to Higgs mass, for which one possible solution is low-energy supersymmetry (SUSY).
Remarkably, it was shown\,\cite{ellis-ross} that the instability of the SM Higgs potential
can be cured without severe fine-tuning by adding bosonic and fermionic
degrees of freedom, ending up with a theory much like SUSY. In the simplest example along this line,
it was also shown~\cite{HX} that adding a new boson-fermion pair does lead to successful
Higgs inflation with a wider range of the tensor-to-scalar ratio
than that of the conventional Higgs inflation.

\vspace*{1mm}

Combining Higgs inflation with SUSY is a challenging task.
For instance, it was found in~\cite{einhorn-jones} that building Higgs inflation in the MSSM
with a minimal K\"ahler potential is not viable~\cite{other}. On the other hand,
in the NMSSM one encounters a tachyonic instability along the direction of the additional singlet scalar~\cite{ferrara1}, though this can be cured by adding higher-order terms
to the K\"ahler potential~\cite{ferrara2}.
Models of this kind were built in the minimal SU(5) GUT with a (strongly) modified
no-scale K\"ahler potential and invoking a large nonminimal Higgs-curvature coupling,
which gave a small tensor-to-scalar ratio $r$ \cite{arai}.
Another type of GUT inflation model is $F$-term hybrid inflation,
which generally leads to very small $r$ as well~\cite{fhi},
though an enhanced value of $r$ could be achieved
with a particular choice of K\"ahler potential~\cite{fhi2}.
Finally, it was shown in~\cite{ns_starobinsky}
that no-scale  supergravity naturally accommodates models of inflation
that yield predictions similar to the Starobinsky model~\cite{Starob}
and the original model of Higgs inflation~\cite{hi}, as well as
allowing more general inflationary potentials that could yield
larger values of $\,r$~\cite{ns_large_r}.

\vspace*{1mm}

In this work, we construct a new class of Higgs inflation models,
using the framework of no-scale supergravity (SUGRA) \cite{su5gut} and
embedded in the supersymmetric SU(5) grand unified theory (GUT).
There are many motivations for this construction, of which we mention three.
Firstly, the inflation scale may well be close to the GUT scale,
according to CMB measurements~\cite{cmb1,cmb2}, so it is very attractive to embed Higgs inflation
into a GUT. Secondly, no-scale SUGRA emerges from simple compactifications of string theory~\cite{witten}.
Thirdly, the flat directions in no-scale SUGRA are advantageous for cosmological applications~\cite{noscale}.

\vspace*{1mm}

For the new type of supersymmetric Higgs inflation in the
no-scale SU(5) GUT framework, we adopt the minimal field content,
with simple extensions of the no-scale K\"ahler potential and minimal SU(5) GUT superpotential.
We derive a generic inflationary potential
that interpolates between a quadratic monomial potential
and a Starobinsky-type potential. The corresponding predictions of
the tensor-to-scalar ratio $\,r\,$ and spectral index $\,n_s^{}\,$ can
accommodate the Planck and BICEP2 observations~\cite{cmb1,cmb2}.
A notable feature of this no-scale SUSY GUT Higgs inflation scenario is that it
does not invoke any non-minimal coupling between the Higgs fields and the Ricci curvature,
i.e., all Higgs bosons couple minimally to gravity via the
energy-momentum tensor. This is an essential difference between our construction
and the conventional SM Higgs inflation~\cite{hi}
as well as its previous SUSY and GUT extensions~\cite{einhorn-jones,ferrara1,ferrara2,arai}.
Finally, we will further analyze the stability of inflation trajectories
in all directions of field space and demonstrate the consistency.


\vspace*{6mm}
\noindent
{\bf 2.\,New Higgs Inflation with No-Scale SUGRA}
\vspace*{3mm}

Our Higgs inflation with no-scale SUGRA and SU(5) GUT contains the following chiral fields
as ingredients: a singlet modulus field $\,T\,$
that may arise from string compactification, a GUT Higgs multiplet $\Sigma$ in the adjoint representation
of SU(5), and a pair of Higgs multiplets $H_1^{}$ and $H_2^{}$ belonging to
$\,\mb{5}$\, and $\,\mb{\bar{5}}\,$ representations of SU(5), respectively.
We postulate the following extended no-scale K\"ahler potential $K$,
which is a hermitian function of these superfields,
 \bge
 \label{Kahler}
 \begin{aligned}
   K \,=\,
   & -3\log\Big[T+T^*-\fr{1}{3}\tr(\Sigma^\dag\Sigma)\\
   & -\fr{1}{3}\Big(|H_1^{}|^2+|H_2^{}|^2-\zeta (H_1^{} H_2^{}+\text{h.c.})\Big)\Big],
   \hspace*{5mm}
 \end{aligned}
 \ede
where  we set the reduced Planck mass $\,\Mp =1\,$ as a mass unit, and
$\,\zeta\,$ is a dimensionless parameter.
The theory obeys a simple $\,\mathbb{Z}_2^{}\,$ symmetry, under which
$\,\Sigma\,$ is odd and all other fields are even.
We also postulate the following holomorphic superpotential $W$\,,
\beqa
   W \,=\, W_\Sigma^{} + W_H^{}\,,
\eeqa
 where the $\,\Sigma\,$ part
\beqa
   W_\Sigma^{} \,=\,
   -\fr{1}{2}m\tr\!\!\(\Sigma^2\)+\fr{1}{4}\tilde\lam\tr\!\!\(\Sigma^4\)
\eeqa
ensures the correct GUT-breaking vacuum for the GUT Higgs $\Sigma$\,,\, and
the $\,H\,$ part
\beqa
W_H^{} \,=\,
\mu H_1^{}H_2^{} - \tilde\be_1^{}H_1^{}\Sigma^2 H_2^{} + \be_2^{}(H_1^{}H_2^{})^2
\label{Sigma2}
\eeqa
generates desired triplet-doublet splitting for the fundamental Higgs multiplets
$H_{1}^{}$\, and $H_{2}^{}$\,.\,
Due to the $\ZZ$ symmetry, the trilinear terms $\,\tr\!(\Sigma^3)\,$ and
$\,H_1^{}\Sigma H_2^{}\,$ are absent in $\,W_\Sigma^{}\,$ and $\,W_H^{}\,$,\,
respectively.

\vspace*{1mm}

When considering spontaneous breaking of the GUT gauge group \,SU(5)\, down to
the SM gauge group $\,\text{SU(3)}\otimes\text{SU(2)}\otimes\text{U(1)}$,\,
the adjoint Higgs field $\,\Sigma\,$ contains the relevant component,
$\,\Sigma \supset $ $\sqrt{2/15\,}\,\text{diag}(1,1,1,-3/2,-3/2)\,\chi$\,,\,
where $\,\chi\,$ is a singlet chiral multiplet, in terms of which
the superpotential $\,W_\Sigma^{}\,$ reduces to
\beqa
W_\Sigma^{} \,=\,
-\fr{1}{2}m\X^2 + \fr{7}{\,120\,}\tilde\lam\,\X^4\,.
\eeqa
The $\,\X\,$ field should have its vacuum expectation value (VEV) take a value
$\,\la \X\ra\equiv u \simeq 2\!\times\! 10^{16}$\,GeV,
as determined by the SUSY GUT gauge unification.
The stationary condition $\,\pd W/\pd \X =0$\,
yields $\,m=\lam u^2$,\,
with $\,\lam \equiv \fr{7}{30}\tilde\lam\,$.

\vspace*{1mm}

In the presence of the GUT breaking VEV $\,u\,$,\,
the Higgs multiplets $H_{1}^{}$ and $H_{2}^{}$ split into
$\,H_1^{}=(H_c,H_u)^T$\, and $\,H_2^{}=(\bar H_c,H_d)^T$\, as usual,
where $\,(H_c,\bar H_c)\,$ are color $SU(3)$ triplets and $\,(H_u,H_d)\,$ are weak
$SU(2)$ doublets. Thus, the $H$ part of superpotential becomes,
\beqa
\label{eq:W-H}
\begin{aligned}
W_H^{} \,=
&~ H_c\big(\mu-\fr{4}{9}\be_1^{} \X^2\big)\bar H_c + H_u\big(\mu-\be_1^{} \X^2\big)H_d
\\
&~ +\be_2^{} (H_c\bar H_c+H_uH_d)^2,
\end{aligned}
\eeqa
where $\,\be_1^{} \equiv \fr{3}{10}\tilde\be_1^{}\,$.\,
In order for the two Higgs doublets $\,(H_u,H_d)\,$ to remain light (at the weak scale)
while the colored Higgs triplets $\,(H_c,\bar H_c)\,$ become heavy,
we set $\,\mu \simeq \be_1^{} u^2$.
The color triplet Higgs bosons then acquire a large mass
$\,M_c = \fr{5}{9}\be_1^{}u^2$\, at tree-level.
We parametrize the Higgs doublets $\,(H_u,\, H_d)$\,
as $\,H_u=(H_u^+,\,H_u^0)^T$\, and $\,H_d=(H_d^0,\,H_d^-)^T$,\,
and identify a linear combination of $\,H_u^0\,$ and $\,H_d^0\,$
as the inflaton. The colored components $\,(H_c,\,\bar H_c)$\,
and the electrically charged components $\,(H_u^+,\,H_d^-)$\,
do not play important r\^oles during inflation,
as we will discuss in Sec.\,4.

\vspace*{1mm}

Examples of modulus stabilisation were studied in~\cite{ns_starobinsky},
and we assume here that non-perturbative ultraviolet dynamics
fixes the VEV of the modulus field to be $\,T=T^*=1/2$\, \cite{TT}.
After these simplifications, the K\"ahler potential and the superpotential become,
\beqs
 \begin{align}
   K\,=\, &-3\log\big[1-\fr{1}{3}\big(|\chi |^2+|H_u^0|^2+|H_d^0|^2\n
   \\[1mm]
          &-\zeta (H_u^0 H_d^0+\text{h.c.})\big)\big],
   \\[2mm]
   W\,=\, &~ \be_1^{} H_u^0\big(u^2-\X^2\big)H_d^0 + \be_2^{}(H_u^0H_d^0)^2\n
   \\[1mm]
          &~ -\fr{1}{2}\lam u^2 \X^2 + \fr{1}{4}\lam \X^4.
 \end{align}
\eeqs
This is the basis for our following analysis of Higgs inflation in the no-scale SUSY SU(5) GUT.


\vspace*{6mm}
\noindent
{\bf 3.~Higgs Inflation Potential and Observables}
\vspace*{3mm}

\begin{figure*}[t]
\centering
\includegraphics[height=7.25cm,width=0.49\textwidth]{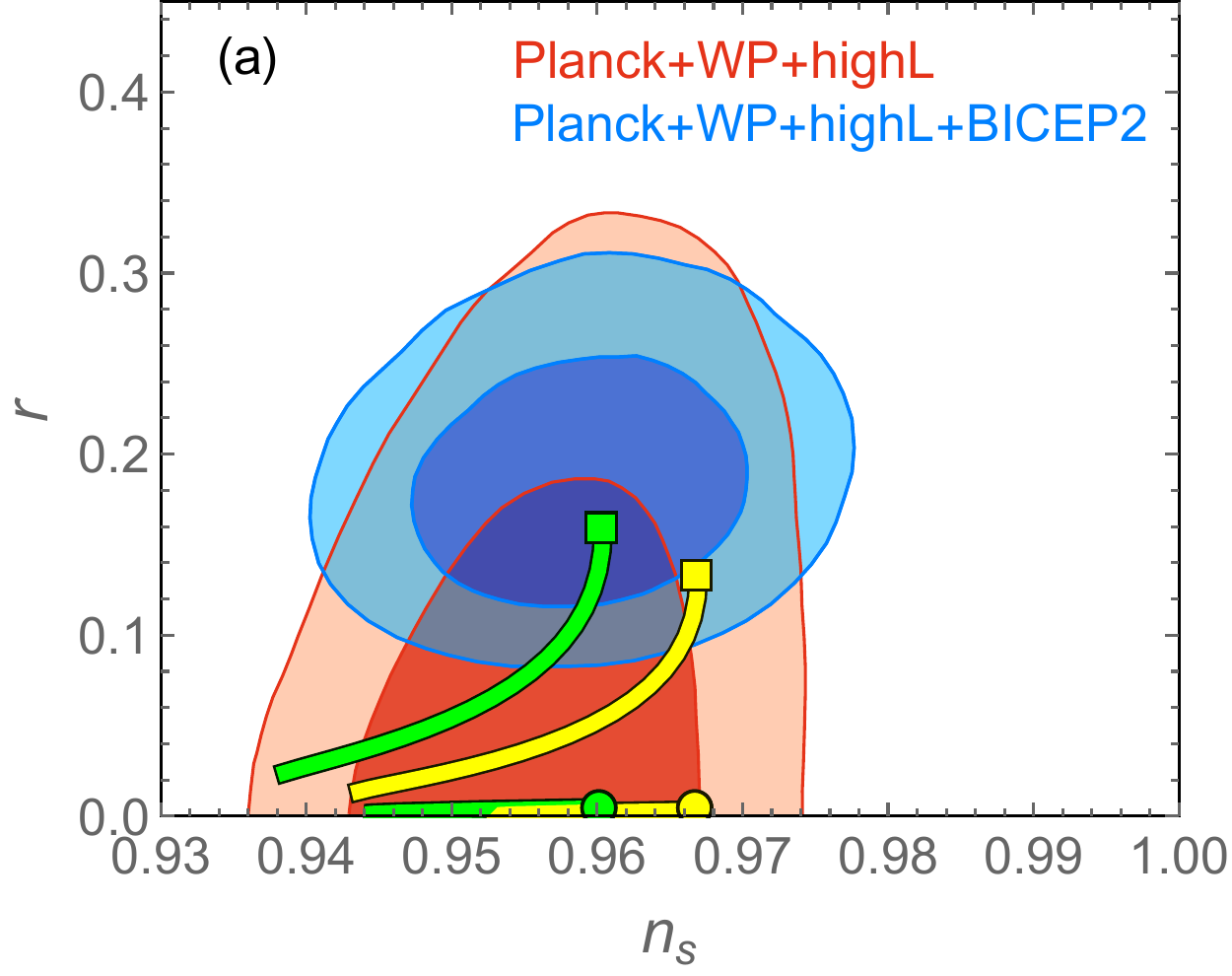}
\includegraphics[height=7.25cm,width=0.49\textwidth]{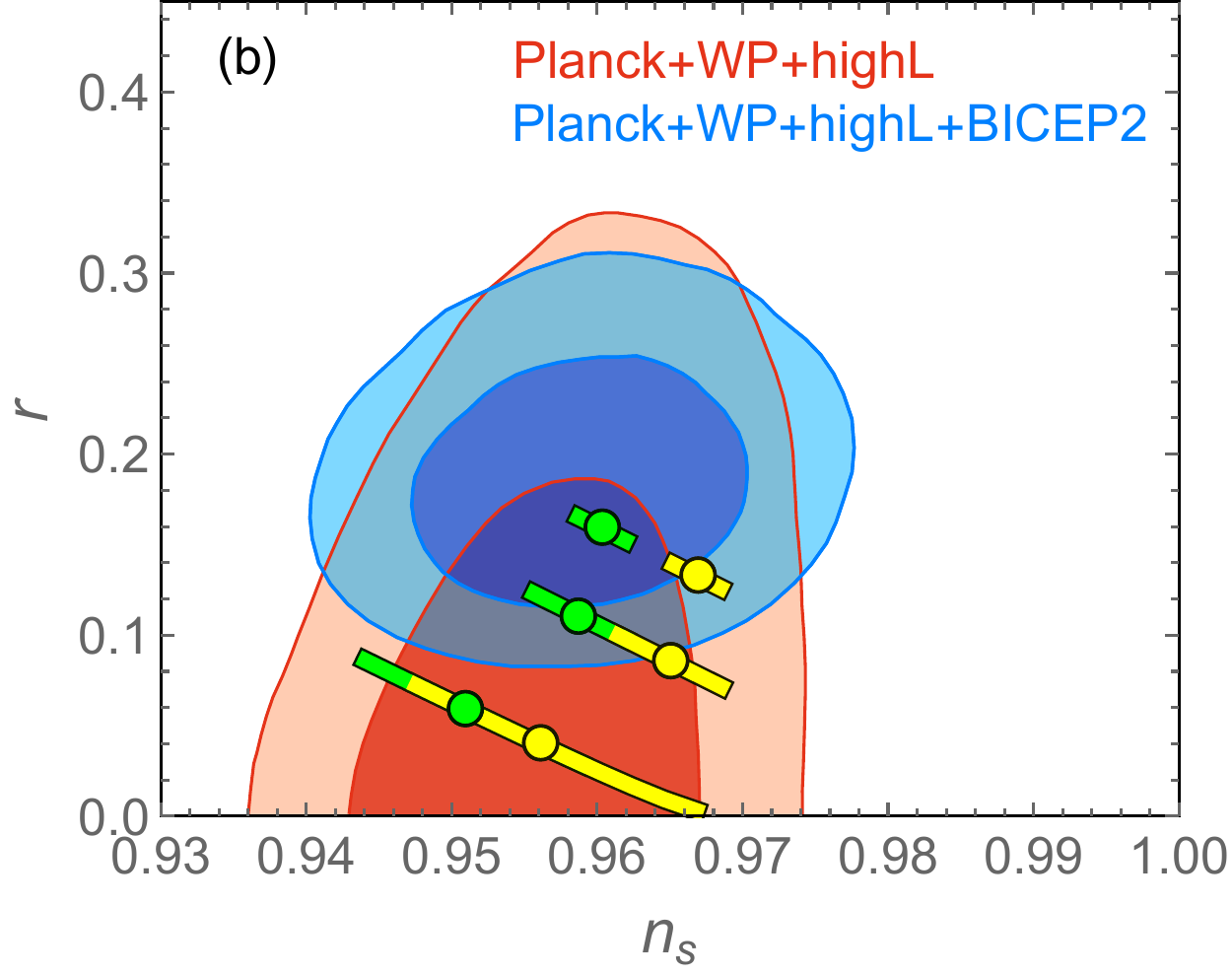}
\caption{The predictions of our no-scale SUSY GUT model of Higgs inflation for the spectral index
$\,n_s^{}$ and tensor-to-scalar ratio $\,r\,$.\,  The green (yellow) dots and curves represent predictions
with 50\,(60) $e$-foldings. In plot (a), the condition $\,\be_2^{}=\fr{1}{3}\be_1^{}(1-\zeta)u^2$\,
is imposed and the round (square) dots correspond to $\,\zeta =0~(\zeta =1)$.\,
The horizontal strips attached to the lower round dots correspond to the effect of varying
$\,\zeta\in [0, 0.1]\,$ (from right to left),
while the upper strips attached to the square dots depict the effect of varying
$\,\zeta\in [0.9, 1]\,$ (from left to right).
In plot (b), the three pairs of dots from top to bottom correspond to
$\,\zeta =(1,\,0.98,\,0.95)$\, and $\,\de=0\,$.\,
The strip attached to each round dot describes the effect of varying $\,\de\,$
over the range $\,\pm \, (1.2 \!\times\! 10^{-3})\,$,\,
via the shifted relation $\,\be_2^{}=\fr{1}{3}\be_1^{}(1-\zeta+\de )u^2$.\,
In each plot, the shaded red and blue (pink and light-blue) contours represent the observational limits at
68\% (\,95\%\,) confidence level as given by Ref.\,\cite{cmb2}.
}
\label{fig:1}
\vspace*{4mm}
\end{figure*}

The $F$-term scalar potential \,$V$\, takes the following standard form,
\beqa
V \,=\, e^{G}\Big( K_{ij^*}^{}\FR{\pd G}{\pd \phi_i^{}}\FR{\pd G}{\pd \phi_j^*}-3\Big) ,
\eeqa
 where $\,G=K+\log W+\log W^*$,\,
 $\,K_{ij^*}$ is the inverse of K\"ahler metric
 $\,K^{ij^*}\equiv \pd^2 K/\pd\phi_i^{}\pd\phi_{j}^{*}$,\,
 and $\,\phi_i^{}\,$ denotes generic scalar fields.
 We parametrize the neutral scalars explicitly as
 $\,H_u^0=\fr{1}{2}(X+Y)e^{\ii\theta}\,$ and
 $\,H_d^0=\fr{1}{2}(X-Y)e^{\ii\varphi}$, which give
 \beqa
 X=|H_u^0|+|H_d^0|\,, ~~~~
 Y=|H_u^0|-|H_d^0|\,. ~~~~
 \eeqa
 The $D$-flatness condition sets $\,Y=0\,$,
 and we further take $\,\theta =\varphi =0$\, in the following
 (the validity of this choice will be examined in Sec.\,4). Thus,
 we can identify the Higgs combination, $\,X=|H_u^0|+|H_d^0|$\,,\, as the inflaton.
 Moreover, during inflation the minimum of $\,\chi\,$ field for a given $\,X\,$
 is always at $\,\chi =0\,$,\, as will be examined in Sec.\,4,
 so we also set $\,\chi=0\,$ from now on.
 Thus, the scalar potential $\,V\,$ becomes a function of $\,X\,$ alone, and
 takes the form
 \beqa
 \label{InfPotentialGeneral}
 V(X)\,=\,
 \frac{\,\be_1^2u^4 \big(1-\fr{\be_2^{}}{\,2\be_1^{}u^2\,}X^2\big)^2X^2\,}
      {2\big(1-\fr{1-\zeta}{6}X^2\big)^2} \, ,
 \eeqa
which is the basis for our subsequent analysis.

\vspace*{1mm}

 The potential (\ref{InfPotentialGeneral}) displays a singularity at $\,X^2=6/(1-\zeta)$\,.\,
 The presence of such a singularity is ubiquitous in no-scale supergravity
 due to the form of the K\"ahler potential.
 It would lead to an exponentially steep potential in terms of
 a canonically-normalised scalar field, violating the slow-roll condition.
 To cure this singularity, we impose the relation
 $\,\be_2^{}=\fr{1}{3}\be_1^{}(1-\zeta ) u^2$,\,
 in which case the scalar potential simplifies to a quadratic monomial,
 \beqa
 \label{InfPotential}
   V(X) \,=\, \fr{1}{2}\be_1^2u^4 X^2 \,.
 \eeqa
This does not lead to a quadratic chaotic inflation model, however,
because the inflaton field $\,X\,$ is not canonically normalised, and
its kinetic term takes the following form
 \beqa
   \ld_{K}(X)\,=\,
   \FR{\,1-\fr{\zeta (1-\zeta )}{6}X^2\,}
      {\,2\big(1-\fr{1-\zeta }{6}X^2\big)^2\,}(\pd_\mu X)^2.
 \eeqa
 Thus, we derive the canonically-normalised inflaton field $\,h$\,
 as a function of $\,X$\,,
 \bge
 \label{CanonField}
 \begin{aligned}
   h \,=&~ \sqrt{6}\,\text{arctanh}
   \FR{(1\!-\!\zeta)X}{\sqrt{6\big(1\!-\!\fr{1}{6}\zeta (1\!-\!\zeta )X^2\big)}}
   \\
        &~ -\sqrt{\FR{6\zeta}{1\!-\!\zeta\,}}\arcsin\!
        \(\!\!\sqrt{\FR{\zeta (1\!-\!\zeta)}{6}}X\!\)\!.
 \end{aligned}
 \ede
From Eq.\,\eqref{CanonField}, we recognize two interesting limiting cases,
$\,\zeta =0$\, and $\,\zeta =1$\,,\,
which can be worked out analytically.

\vspace*{1mm}

When $\,\zeta =0$\,,\, we have the simplified relation
$\,X(h)=\sqrt{6}\,\tanh(h/\sqrt 6)$,\, and the potential $\,V(h)\,$ becomes,
 \bge
 V(h) \,=\, 3\be_1^2u^4\tanh^2\!\FR{h}{\sqrt{6}\,} \,,
 \ede
 which is exponentially flat at large $\,h\,$,\,
 and analogous to the Starobinsky model~\cite{Starob}.
 In this case, we can derive the following predictions
 for the primary inflationary observables,
 $\,(n_s^{},\,r) \simeq (0.967,\,0.003)$\, for 60 $e$-foldings,
 which are similar to the predictions of the Starobinsky model~\cite{Starob}
 and conventional Higgs inflation~\cite{hi}, as expected.

\vspace*{1mm}

 On the other hand, when $\,\zeta =1\,$,\, the $\,X\,$ field is already normalised canonically,
 $\,X(h)=h\,$,\, and we recover a quadratic monomial potential.
 This limit therefore yields the same predictions as quadratic chaotic inflation,
 namely, $\,(n_s^{},\,r) \simeq (0.967,\,0.130)$\, for 60 $e$-foldings.

\vspace*{1mm}

 When $\,\zeta\,$ varies between \,0\, and \,1\,,\, we obtain a class of inflation potentials
 that extrapolate between the quadratic monomial and Starobinsky-type potentials.
 The predictions for $\,(n_s^{},\,r)$\, can be worked out numerically.
 We plot them in Fig.\,\ref{fig:1}(a), where the green (yellow) dots and their attached curves
 represent predictions with 50\,(60) e-foldings.
 In this plot, the round (square) dots correspond to $\,\zeta =0~(\zeta=1)\,$.\,
 The horizontal strips attached to the lower round-dots correspond to
 the effect of varying $\,\zeta\in [0, 0.1]\,$ (from right to left),
 while the upper strips attached to the square-dots depict the effect of varying
 $\,\zeta\in [0.9, 1]\,$ (from left to right).
 For comparison, we have depicted the recent observational limits~\cite{cmb1,cmb2}
 as the shaded red and blue (pink and light-blue) contours at
 68\% (\,95\%\,) confidence level.
 We see that our predictions for $\,(n_s^{},\,r)$\,
 are well compatible with the current data.

\vspace*{1mm}

 For Fig.\,\ref{fig:1}(a), we imposed the constraint
 $\,\be_2^{}=\fr{1}{3}\be_1^{}(1-\zeta )u^2$,\, and
 it is interesting to check what happens when $\,\be_2^{}\,$
 deviates slightly from this relation.
 For this purpose, our starting point will be (\ref{InfPotentialGeneral}) and (\ref{CanonField}).
 We find that shifting $\,\be_2^{}\,$ according to
 $\,\be_2^{}=\fr{1}{3}\be_1^{}(1-\zeta +\de )u^2$,\,
 with $\,\de\,$ varying over the range $\,\pm \, (1.2 \!\times\! 10^{-3})$,\,
 will make the predicted \,$(n_s^{},\, r)$\, values vary
 as shown by the green and yellow strips (attached to the corresponding dots)
 in Fig.\,\ref{fig:1}(b). Here the three pairs of round dots from top
 to bottom correspond to $\,\zeta =(1,\,0.98,\,0.95)$,\, while
 the green (yellow) dots and strips represent the predictions with 50\,(60) e-foldings.
 Fig.\,\ref{fig:1}(b) shows again that the predicted values of $\,(n_s^{},\,r)$\,
 agree well with the current experimental limits \cite{cmb1,cmb2}.
 Together with the minimal case (\ref{InfPotential}) presented in Fig.\,\ref{fig:1}(a),
 Fig.\,\ref{fig:1}(b) shows that our model predicts the range of the tensor-to-scalar ratio $\,r\,$
 to be in the range $\,\order{10^{-3}\!-\!10^{-1}}$,\,
 without contrived theoretical inputs~\cite{foot-1}.

\vspace*{1mm}

 The coefficient $\,\be_1^{}\,$ in the potential (\ref{InfPotential})
 is fixed by the amplitude of scalar perturbations,
 $\,(V/\ep)^{1/4}\simeq 0.027$\,,\, as measured by the Planck satellite \cite{cmb1}.
 This yields $\,\be_1^{}\simeq 0.06$\, with little variation when $\,\zeta\,$ changes
 between \,0\, and \,1\,.\,  Taking the GUT breaking scale $\,u\simeq 0.01\,$ in units of the reduced
 Planck mass $\,\Mp \simeq 2.44\times 10^{18}\,$GeV,
 we estimate the mass of colored triplet Higgs bosons to be $\,M_c\simeq 0.8\times 10^{13}\,$GeV.
 There are arguments\,\cite{GotoNihei} from proton stability
 and gauge coupling unification
 that prefer $M_c$ to be much larger than $10^{13}$\,GeV~\cite{foot-3}.
 However, the proton stability argument relies on particular hypotheses about the mechanism
 for generating the light fermion masses and mixing in the SU(5) GUT.
 This is a known problem of the model and can be evaded in various ways
 (which are beyond the scope of this short paper).
 Also, the unification argument depends on over-simplified assumptions about physics around
 and beyond the GUT scale that may be relaxed without affect our main predictions.
 For example, the non-minimal contributions to the gauge kinetic function in supergravity
 will modify the gauge unification condition.

\vspace*{1mm}

 Finally, we note that the $\beta_2^{}$-term in the superpotential \eqref{eq:W-H}
 induces a new dimension-4 term (involving the light Higgs doublets $H_u$ and $H_d$)
 in the Higgs potential. However, the coefficient of this term is only of the order
 of $\,\bar{\mu}\beta_2^{}\simeq \frac{1}{3}\beta_1^{}(1\!-\!\zeta)u^2\bar{\mu}\,$
 in Planck mass units (or
 $\,\bar{\mu}\beta_2^{}\simeq 0.02(1-\zeta)u^2\bar{\mu}/M_{\text{Pl}}^3 \lll 1\,$
 with the Planck-mass dependence exhibited),
 where $\,\bar{\mu}=\mu-\beta_1^{}u^2=\order{\text{TeV}}\,$
 is the residual $\mu$-term at the electroweak scale. Hence, this new term is negligible
 for low-energy SUSY phenomenology. Other induced higher-dimensional operators in
 the Higgs potential are even more suppressed by powers of $\,1/\Mp\,$.


\vspace*{6mm}
\noindent
{\bf 4.~Stability of the Inflationary Trajectory}
\vspace*{3mm}

In this section, we study the stability of inflationary trajectory in our Higgs inflation scenario.
Checking stability is a necessary and nontrivial task
since some previous proposals for SUSY Higgs inflation
suffer from tachyonic instabilities, as mentioned in Sec.\,1.
For this purpose, we compute the effective mass matrix
$\,M_{ij}^{}\equiv\fr{1}{2}(\pd^2V/\pd\phi_i^{}\pd\phi_j^{})$\,
in all scalar directions $\,\phi_i^{}\,$ along the inflationary trajectory
for $\,X <\sqrt{6/(1\!-\!\zeta)\,}\,$ (with $\,\zeta<1$),\,
and all other fields $\,\phi_j^{}=0$\,.\,
The mass matrix is block-diagonal, and the analysis can be subdivided into four independent sectors,
namely, the colored Higgses $(H_c,\,\bar H_c)$,\, the charged Higgses $(H_u^\pm,\,H_d^\pm)$,\,
the phases of neutral Higgses $(\theta,\,\varphi)$,\,
and the two components $(s,\,t)$ of the SM gauge-singlet scalar $\,\X=s+it$\,.

\vspace*{1mm}

 In the case of the color-triplet Higgs fields $(H_c,\,\bar H_c)$,
 we consider the first color component of each field for simplicity, which we parametrize as
 $\,h_c e^{\ii\theta_c}\,$ and $\,\bar h_c e^{\ii\bar\theta_c}\,$.\,
 Steepness along the $\,h_c-\bar h_c\,$ direction is guaranteed by the $D$-term in the effective potential,
 and the effective mass along the $\,h_c+\bar h_c\,$ direction is
 $\,M_{cc}^2=2\be_1^2u^4$\, \cite{foot-2}.
 We have also checked that there are no instabilities  in the $(\theta_c,\,\bar\theta_c)$ directions.
 Therefore, there is no instability in this sector.
 The check for the charged Higgs sector is similar.

 Regarding the angular parts of neutral Higgs bosons $(\theta,\,\varphi)$,\,
 we find the following elements in the $2\!\times\!2$ effective mass matrix $\,{\cal M}^2\,$,
 \beqs
 \beqa
 \hspace*{-5mm}
 && M_{\theta\theta}^2 \,=\, M_{\varphi\varphi}^2
 \nn\\[1.5mm]
  \hspace*{-5mm}
 && =\, \FR{\be_1^2u^4X^4}{\,\big(1\!-\!\frac{1-\zeta}{6}X^2\big)^{\!4}\,}
     \left[\,\frac{1}{3}+\frac{\,8\zeta^2\!+\!17\zeta \!-\!7}{144}X^2 \right.
 \\[1.5mm]
 \hspace*{-5mm}
 && \left.
 ~~~+\frac{\,5\zeta^3\!-\!6\zeta\!+\!1\,}{432}X^4
    +\frac{\,(1\!-\!\zeta)^2(17\zeta^2\!+\!4\zeta\!-\!3)}{20736}X^6\right]\!,
 \hspace*{5mm}
 \nn
 \\[4mm]
 \hspace*{-5mm}
 && M_{\theta\varphi}^2=M_{\varphi\theta}^2
 \nn\\[1.5mm]
 \hspace*{-5mm}
 && =\, \FR{\be_1^2u^4X^4}{\big(1\!-\!\fr{1-\zeta}{6}X^2\big)^{\!4}}
        \left[\frac{\,3\zeta\!+\!1\,}{12}+\frac{\,14\zeta^2\!+\!5\zeta\!-\!1\,}{144}X^2
        \right.
\\[1.5mm]
 \hspace*{-5mm}
 && \left.
   ~~~+\frac{\,13\zeta^3\!-\!9\zeta^2\!-\!3\zeta\!-\!1\,}{864}X^4
      +\frac{\,\zeta(10\zeta\!-\!1)(\zeta\!-\!1)^2}{10368}X^6\right]\!.
 \nn
 \eeqa
 \eeqs
 The two eigenvalues of $\,{\cal M}^2\,$ are
 \beqs
 \beqa
 \hspace*{-4mm}
  &&
  M_1^2\,=\,\FR{\be_1^2u^4X^4}{\,\big(1\!-\!\fr{1-\chi}{6}X^2\big)^{\!4}\,}
  \left[\frac{\,3\zeta\!+\!5\,}{12}+\frac{\,11\zeta^2\!+\!11\zeta\!-\!4\,}{72}X^2\right.
  \nn\\[1.5mm]
  \hspace*{-4mm}
  && \left.
  +\frac{\,23\zeta^3\!-\!9\zeta^2\!-\!15\zeta\!+\!1\,}{864}X^4
  +\frac{\,(1\!-\!\zeta)^2(37\zeta^2\!+\!2\zeta\!-\!3)\,}{20736}X^6\right]\!,~~
  \nn\\
  \hspace*{-4mm}
  &&
  \\
 \hspace*{-4mm}
  &&
  M_2^2 \,=\,
  \FR{\be_1^2u^4X^4}{\,\big(\!1-\frac{1-\zeta}{6}X^2\big)^{\!4}}
  \left[\frac{\,1\!-\!\zeta\,}{4}-\frac{\,(1\!-\!\zeta)^2\,}{24}X^2\right.
  \nn\\[1.5mm]
  \hspace*{-4mm}
  &&
  \hspace*{9mm}\left.
  ~+\frac{\,(1\!-\!\zeta)^3\,}{288}X^4-\frac{\,(1\!-\!\zeta)^4\,}{6912}X^6\right]\!,
 \eeqa
 \eeqs
which are both positive for $\,\zeta\in [0,\,1]\,$
 and $\,X <\sqrt{6/(1\!-\!\zeta)}$, during the inflationary epoch.
 Hence, the effective mass matrix
 $\,{\cal M}^2\,$ is positive definite along the inflation trajectory.

\vspace*{1mm}

Finally, we compute the effective mass matrix for the real and imaginary parts
 of the singlet field $\,\X\,$, up to corrections of $\,\order{u^2}$\,,
 \bge
   M_{ss}^2 \,=\, M_{tt}^2 \,=
   \FR{\be_1^2X^4}{\,4\big(1\!-\!\frac{\,1-\zeta\,}{6}X^2\big)^2\,} \,,
   ~~~~
   M_{st}^2\,=\,0\,.~~~~
 \ede
 %
 This is always positive-definite, so the $(s,t)$ directions are also stable.
 In summary, we have systematically verified that the inflationary trajectory is
 stable in all scalar directions for typical parameter choices.


\noindent
{\bf 5.~Conclusions}
\vspace*{3mm}

Higgs inflation provides a highly economical and predictive approach for the
cosmic inflationary paradigm. In this work, we have proposed a new class of
Higgs inflation models in the framework of an SU(5) GUT embedded in no-scale supergravity.
The structure of this type of models is fairly simple,
since it includes a near-minimal no-scale K\"ahler potential
and simple superpotential with terms up to fourth order in the Higgs chiral multiplets.
The resultant inflaton potential has a variable form, capable of interpolating
between quadratic monomial and Starobinsky-type potentials.
These models can therefore accommodate a wide range of values of the tensor-to-scalar ratio
$\,r$\,,\, while predicting values of the scalar spectral index $\,n_s^{}\,$
that are compatible with the present experimental limits.
Future CMB observations will soon measure or constrain
more precisely the possible value of $\,r\,$.\,
These will further test the predictions of this new class of Higgs inflation models
with no-scale SUSY GUT.

\vspace*{4mm}
\noindent
{\bf Acknowledgements}
\\[2mm]
The work of JE was supported in part by the London Centre for Terauniverse
Studies (LCTS), using funding from the European Research Council
via the Advanced Investigator Grant 267352, and in part by the STFC Grant ST/J002798/1.
HJH and ZZX were supported by Chinese NSF (Nos.\ 11275101 and 11135003)
and National Basic Research Program (No.\ 2010CB833000).


\begin{thebibliography}{99}
\bibitem{foot-1}
 We do not discuss here two- or multi-field effects \cite{KK},
 which would tend to suppress $\,r\,$.
\bibitem{foot-3}
 If, instead of the last two quartic terms in the superpotential $W_H^{}$ of (\ref{Sigma2}),
 we had used a more conventional trilinear term
 $\,\epsilon H_1 \Sigma H_2$,\, the amplitude of scalar perturbations would have required
 $\,\epsilon \sim 10^{-6}$,\, which would have implied $\,M_c\simeq 10^{10}\,$GeV.
 Avoiding such a rather low mass $M_c$ was in fact our principal motivation for imposing
 the $\ZZ$ symmetry discussed earlier.
\bibitem{foot-2}
 The effective mass $\,M_{cc}$ calculated here should not be
 confused with the color-triplet Higgs mass $\,M_c\,$ computed earlier.
 The mass $\,M_c\,$ was derived in the GUT-breaking vacuum, whereas
 $\,M_{cc}\,$ is evaluated at $\,\X=0$\,,\, and is valid for $\,X\sim \order{1}$.
\end{thebibliography}
\end{document}